\documentclass[sigconf]{acmart}

\AtBeginDocument{%
  }

\copyrightyear{2026}
\acmYear{2026}
\setcopyright{cc}
\setcctype{by}
\acmConference[CAIS '26]{ACM Conference on AI and Agentic Systems}{May 26--29, 2026}{San Jose, CA, USA}
\acmBooktitle{ACM Conference on AI and Agentic Systems (CAIS '26), May 26--29, 2026, San Jose, CA, USA}
\acmDOI{10.1145/3786335.3813201}
\acmISBN{979-8-4007-2415-2/2026/05}

\usepackage{booktabs}
\usepackage{tabularx}
\usepackage{caption}
\usepackage{xcolor}
\definecolor{diffadd}{RGB}{0, 150, 0}
\definecolor{diffdel}{RGB}{200, 0, 0}
\newcommand{\add}[1]{\textcolor{diffadd}{#1}}

\newsavebox{\delbox}
\newcommand{\del}[1]{%
  \begingroup
  \savebox{\delbox}{\textcolor{diffdel}{#1}}%
  \rlap{\usebox{\delbox}}%
  \textcolor{diffdel}{\rule[0.5ex]{\wd\delbox}{0.4pt}}%
  \endgroup
}

\definecolor{darkblue}{rgb}{0.0, 0.0, 0.55}
\definecolor{darkgray}{rgb}{0.2, 0.2, 0.2}
\newcommand{\testdesc}[1]{\textcolor{darkgray}{\texttt{\textit{\breakaliases{#1}}}}}
\makeatletter
\newcommand{\breakaliases}[1]{\@tfor\@ch:=#1\do{%
  \if\@ch-\@ch\allowbreak\else\if\@ch\_\@ch\allowbreak\else\@ch\fi\fi}}
\makeatother

\usepackage{listings}

\begin{document}

\title{Scaling Expert Feedback with Reflective Edit Propagation in Compositional Knowledge Bases}


\newcommand{\customaffiliation}[4]{
  \affiliation{
    \institution{#1}
    \city{#2}
    \state{#3}
    \country{#4}
  }
}

\author{Jiajing Guo}
\email{jiajing.guo@us.bosch.com}
\customaffiliation{Bosch Research North America}{Sunnyvale}{California}{USA}

\author{Xueming Li}
\email{xueming.li@de.bosch.com}
\customaffiliation{Robert Bosch GmbH}{Reutlingen}{Baden-Württemberg}{Germany}

\author{Jorge Piazentin Ono}
\email{jorge.piazentinono@us.bosch.com}
\customaffiliation{Bosch Research North America}{Sunnyvale}{California}{USA}

\author{Wenbin He}
\email{wenbin.he2@us.bosch.com}
\customaffiliation{Bosch Research North America}{Sunnyvale}{California}{USA}

\author{Liu Ren}
\email{liu.ren@us.bosch.com}
\customaffiliation{Bosch Research North America}{Sunnyvale}{California}{USA}

\renewcommand{\shortauthors}{Guo et al.}

\begin{abstract}
Domain-specific knowledge bases (KBs) encode vertical expertise and proprietary information that organizations depend on, but curating them at scale is a persistent challenge. Although Large Language Models (LLMs) can draft initial entries efficiently, technical accuracy still requires human expert validation, and reviewing entries one by one at scale is impractical. We present \textbf{R}eflective \textbf{A}gent for \textbf{I}dentifier \textbf{D}ictionary (\textbf{RAID}), a novel system that transforms individual expert edits into systematic knowledge updates. Unlike traditional ``correct-and-save'' paradigms, RAID utilizes a reflective agent to infer the underlying semantic intent behind a single expert edit and propagates that correction across the entire KB through a three-step architecture: Intent Inference, Reflection-based Planning, and User Controlled Execution. We evaluated the reflection and propagation performance on a public dataset and conducted a user study with subject matter experts with proprietary data. The evaluation shows RAID's technical feasibility in capturing expert intent and its potential to scale specialized expertise across industrial knowledge bases
\footnote{Demo materials: \href{https://ripple-brass-634.notion.site/RAID-CAIS-26-Demo-Materials-34fc7f157b288088a770dda9108a3f43}{project page link}.}
.
\end{abstract}





\maketitle

\section{Introduction}

\begin{figure*}[t]
    \centering
    \includegraphics[width=\textwidth]{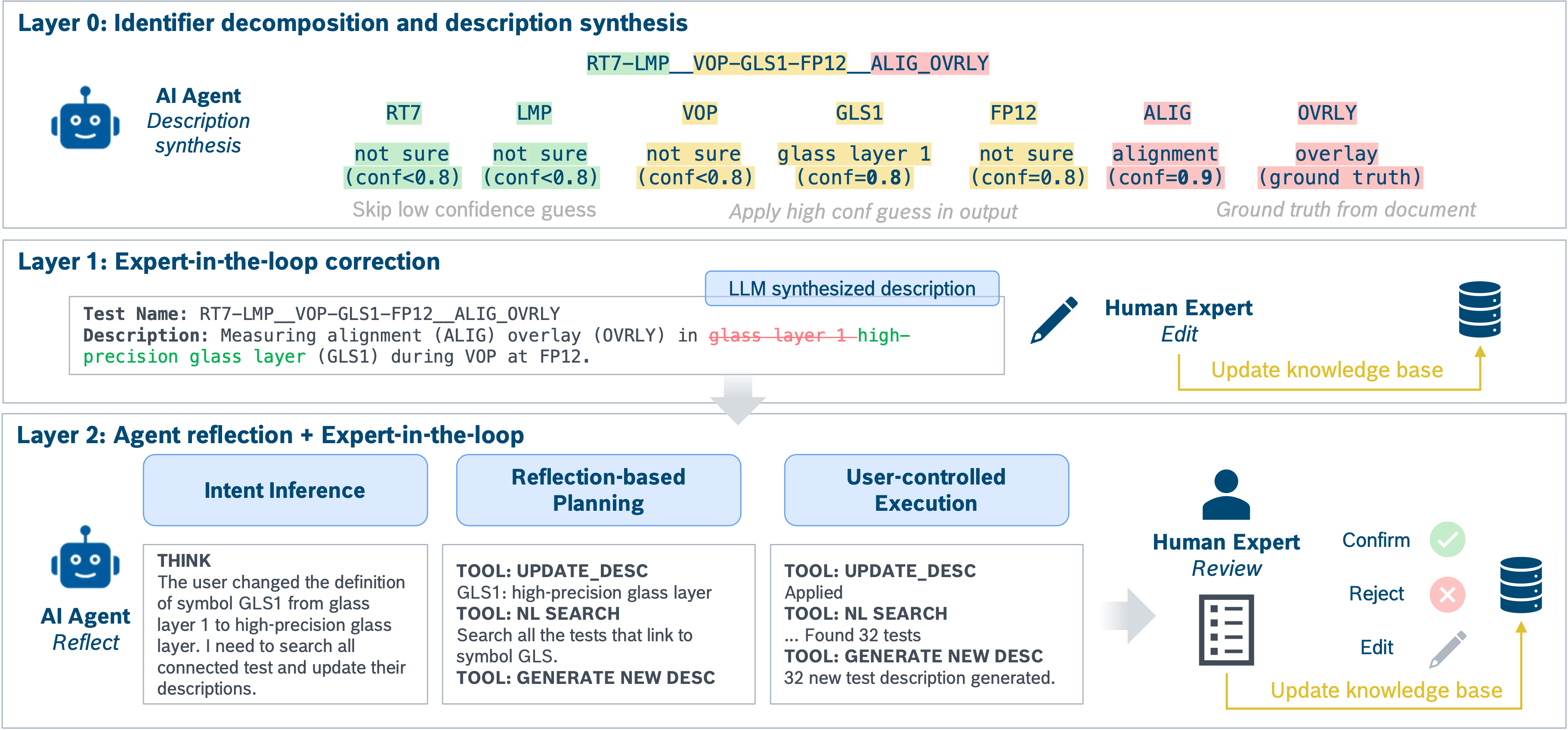}
    \caption{Reflective agent framework. 
    Layer 0 represents the LLM synthesis phase where an LLM decomposes identifiers (semiconductor test names as an example) based on specific patterns, creates hierarchical relationships, and synthesizes test descriptions. 
    Layer 1 represents the expert-in-the-loop validation phase where the human expert makes edits and updates the knowledge base. 
    Layer 2 represents the reflective agent framework where an LLM agent infers the expert’s intent, creates a chain-of-tools plan to propagate the changes, and executes them based on user control. The expert needs to review the propagated results to choose to accept, reject, or edit the results.
    }
    \label{fig:framework}
\end{figure*}

Domain-specific knowledge bases (KBs) support many vertical workflows in organizations~\cite{Anjelia2025-Agent-Organizational}, and Large Language Models (LLMs) have made it efficient to draft initial entries at scale.
The bottleneck has shifted from authoring to validation: in vertical domains, LLM-generated content frequently contains technical inaccuracies~\cite{Szymanski2025-Judge}, so expert review is mandatory, but reviewing entries one by one does not scale to dictionaries with tens of thousands of entries.

This problem is especially acute for \textit{compositional identifiers}, structured strings assembled from reusable symbols that organizations adopt for standardization.
Examples span domains: LOINC composes laboratory test codes from six axes (Component, Property, Time, System, Scale, Method); RxNorm builds clinical drug concepts from Ingredient, Dose Form, and Strength; SWIFT/BIC codes in finance use four segments; InChIKey hashes index chemical compounds; and in manufacturing, parameter identifiers encode production line monitoring configurations.
While many of these have public references, in organizational settings such identifiers are often \textbf{poorly documented}, and much of the meaning lives in the heads of the engineers who designed the naming conventions.
Building a KB that decodes these identifiers, what we call an \textit{identifier dictionary}, is therefore a pressing need.

The compositional structure that makes these dictionaries useful also defines the validation challenge: a correction to one symbol's meaning should propagate to every entry that contains it.
Existing approaches address similar issues from several perspectives.
LLM-enhanced KG curation tools support human-in-the-loop validation of individual entries~\cite{Bikaun2024-CleanGraph, Tsaneva2025-KG-Validation} but treat each correction as local.
Agent memory systems let agents self-edit an external store to accumulate experience~\cite{Packer2023-MemGPT, Park2023-Generative, Zhao2024-ExpeL}, but the agent is the principal and there is no human-validation step.
Parametric knowledge editing~\cite{Cohen2024-yp, Zhao2024-RIPPLECOT} studies how factual updates ripple within model weights, not within an external KB.
Little work directly addresses what an organizational curation workflow needs: inferring expert intent from a single edit and cascading the corresponding correction across structurally related entries in a shared KB, under expert control.
We frame this as a single research question: \textbf{How can we efficiently utilize limited human feedback to maintain a massive compositional knowledge base?}

Answering this requires reasoning that goes beyond find-and-replace.
The system should (1) \textbf{generalize} beyond the literal edit: if an expert corrects one ingredient code in RxTerms, related ingredients or dose forms may need analogous updates that a programmatic rule cannot infer; and (2) \textbf{reason backwards} from the identifier level, since experts typically edit assembled descriptions rather than individual symbols, and the system should determine which underlying symbol the edit actually targets.

We present \textbf{RAID} (\textbf{R}eflective \textbf{A}gent for \textbf{I}dentifier \textbf{D}ictionary), a system that uses a single expert edit as a reflection trigger, infers the expert's latent intent (e.g., updating a specific symbol's meaning), and proposes propagation across the dictionary under expert control.
Our contributions are:
\begin{enumerate}
    \item A \textbf{reflective agent architecture} that infers expert intent from a single edit and propagates corrections across compositional identifier structures via chain-of-tools planning.
    \item An \textbf{interactive interface} that lets domain experts review, accept, or refine the agent's proposed changes in the loop.
    \item A quantitative benchmark on a public dataset with synthesized human edits and a user study with experts that demonstrate technical feasibility in an industrial setting.
\end{enumerate}

\begin{figure*}[t]
    \centering
    \includegraphics[width=\linewidth]{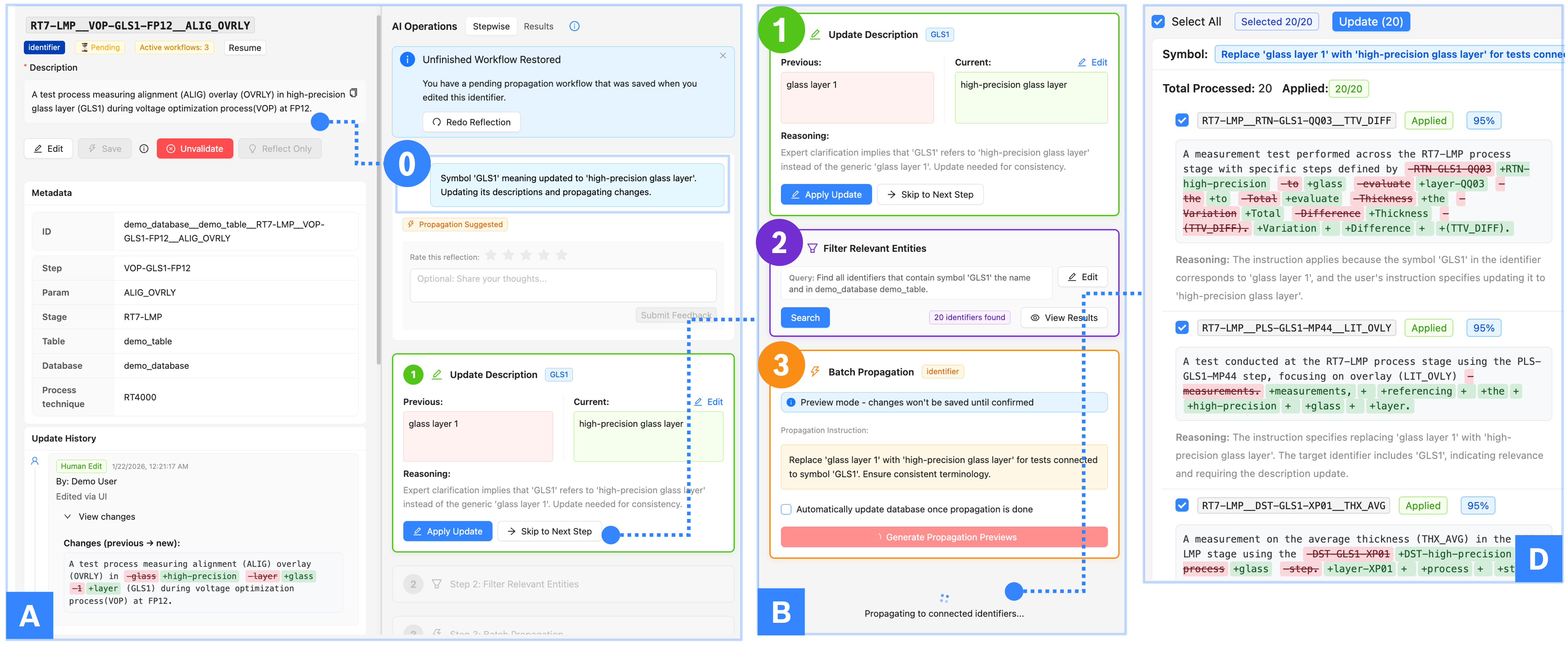}
    \caption{
    RAID system and interaction flow. 
    The primary interactions occur in the \textit{Propagation Panel} (\textbf{B}) within the \textit{Detail View} (\textbf{A}) page.
    After a user edits and saves a description, the panel displays a message summarizing the inferred intent \textcircled{0} and the agent-proposed plan sequences, displayed as individual cards.
    The \textit{Update Card} (\textcircled{1}) allows the user to validate the proposed symbol description update.
    The \textit{Search Card} \textcircled{2} explains how the agent will search for tests to apply the propagation.
    The \textit{Revision Card} \textcircled{3} explains how the agent will update descriptions for the tests retrieved in Step \textcircled{2}.
    The Propagation panel (\textbf{B}) demonstrates the stepwise interaction, where Step \textcircled{2} and \textcircled{3} are grayed out before the user completes Step \textcircled{1}.
    The propagation results are displayed under \textit{Revision Card} (\textbf{D}), awaiting expert approval.}
    \label{fig:raid_ui}
\end{figure*}

\section{RAID: Reflective Agent System for Identifier Dictionary}

We present \textbf{RAID} (\textbf{R}eflective \textbf{A}gent System for \textbf{I}dentifier \textbf{D}ictionary) for the curation of domain-specific knowledge bases. 
In this section, we first present the algorithm of the framework, followed by an example scenario of expert revision.

\subsection{System Architecture and Algorithm}
The core architecture is built upon a three-stage loop that mimics the cognitive process of a human expert (Figure~\ref{fig:framework}). When a user edits a single entry, which can be a symbol or an identifier, the agent initiates the following sequence.

\textbf{Intent Inference}.
After a user modifies an identifier or symbol entry, the Reflection Agent compares the original text and corrected text to infer the underlying semantic intent.
We summarized three types of corrections based on a formative study:
\textbf{semantic confusion corrections}, where a symbol's meaning is corrected from a related but incorrect definition (e.g., \testdesc{GLS1} corrected from ``glass layer 1'' to ``high-precision glass layer'');
\textbf{granularity corrections}, where a vague description is replaced with one capturing distinguishing technical details (e.g., ``a blood pressure medication'' replaced with ``a dihydropyridine calcium channel blocker that relaxes vascular smooth muscle'');
and \textbf{surface edits}, such as case normalization or abbreviation expansion that does not change meaning and should not trigger propagation.
We use the example edits collected in the formative study as few-shot examples in the prompt, and instruct the agent to determine whether the edit represents a systemic update to a symbol's definition or a minor fix.

\textbf{Reflection-based Planning}.
The agent generates a sequence of actions (Chain-of-tools) in JSON format. The system is equipped with three tools: \textit{Update Description}, \textit{NL Search}, and \textit{Generate New Description}. Descriptions and examples are shown in Table~\ref{tab:tools}.

\textbf{User Controlled Execution}.
Following the suggested practice in human-AI collaboration~\cite{Shneiderman2020-HAI}, human experts are the final approvers. The system does not apply changes immediately after proposing the plan. Instead, the new description candidates are presented on the interface and await expert validation.

\subsection{Demo Scenario}

A subject matter expert wants to validate the LLM-synthesized test (identifier) descriptions. The user starts from the dictionary dashboard, where they browse entries and select a test identifier (e.g., \testdesc{SA33-TPL\_\_XREG-EVST-F01\_\_OVSP\_LAGN}) to open its \textit{Detail View}. The user then corrects the description:
\texttt{A test for \del{cross-reference register} \add{cross routing register} (XREG) stage voltage overshoot (OVSP) in LAGN pattern 01}.

Instead of a simple save, the edit triggers the reflective agent. The \textit{Propagation Panel} appears with a sequence of cards representing the agent's proposed plan. First, a \textit{Message Card} explains the agent's reflection (e.g., ``The user redefined \testdesc{XREG} as \testdesc{cross routing register}''). Next, an \textit{Update Card} presents the inferred symbol update with old and new descriptions side by side. The user can review and edit the proposed update before clicking \textit{Apply Update} to proceed.

The interaction follows a stepwise design inspired by Cognitive Forcing Functions~\cite{Bucinca2021-Cognitive-Forcing}, where each step of the plan requires explicit validation before the next step becomes active. After approving the symbol update, a \textit{Search Card} is enabled, showing the natural language query that will filter relevant identifiers for propagation. The user can modify this query and then trigger the search. Finally, a \textit{Revision Card} displays the instruction used to generate updated descriptions for all matched entries. The user clicks \textit{Generate Propagation Preview} to see the batch results and can accept, reject, or edit individual entries.


\subsection{Implementation}
The system is implemented using React for the frontend, Django and PostgreSQL for the backend and database. The LLM implementation uses the Langchain package and GPT-5-mini for reflection and planning. 
We selected GPT-5-mini to balance reasoning quality against the latency and cost requirements of an interactive workflow, where each edit triggers multiple LLM calls.

\begin{table*}[htbp]
  \centering
  \small
  \caption{Chain-of-tools configurations. 
  The agent is provided with the following tool configurations and is tasked with planning a sequence of actions by populating the required parameters. 
  }
  \label{tab:tools}
  \setlength{\tabcolsep}{6pt} 
  \renewcommand{\arraystretch}{1.3} 
  \begin{tabularx}{\textwidth}{ 
    >{\hsize=0.18\hsize\raggedright\arraybackslash}X 
    >{\hsize=0.22\hsize\raggedright\arraybackslash}X 
    >{\hsize=0.30\hsize\raggedright\arraybackslash}X 
    >{\hsize=0.30\hsize\raggedright\arraybackslash}X 
    }
    \toprule
    \textbf{Tool} 
    & \textbf{Description} 
    & \textbf{Parameters} 
    & \textbf{Example Use} \\
    \midrule
    Update Description 
    & Updates the existing description of a symbol or identifier.
    & \texttt{name: symbol or test name} \newline 
      \texttt{old\_desc: old description} \newline 
      \texttt{new\_desc: new description} 
    & \texttt{name: GLS1} \newline 
      \texttt{old\_desc: glass layer 1} \newline 
      \texttt{new\_desc: high-precision glass} \\
    \midrule
    NL Search 
    & Generates and executes SQL to return relevant candidates
    & \texttt{nl\_query: natural language query to instruct the SQL agent to generate SQL} 
    & \texttt{nl\_query: Search all the tests that connect to symbol ``GLS1''} \\
    \midrule
    Generate New Description 
    & Generates new descriptions of the retrieved candidates in \textit{NL Search} step
    & \texttt{nl\_inst: natural language instruction to revise the symbol or test}
    & \texttt{nl\_inst: Replace ``glass layer 1'' with  ``high-precision glass''} \\
    \bottomrule
  \end{tabularx}
\end{table*}

\section{Evaluation}

We conducted two experiments to evaluate the technical performance and user experience of our system.

\subsection{Quantitative Evaluation}
\label{sec:quant-eval}

We evaluated RAID on a benchmark constructed from RxTerms~\cite{RxTerms}, a publicly released drug interface terminology derived from RxNorm.
RxTerms shares the compositional structure of our primary case study: each drug identifier (e.g., \textit{amlodipine 5\,MG / hydrochlorothiazide 12.5\,MG Oral Tablet}) is assembled from reusable symbols denoting ingredients, dose forms, and routes.
This makes it a suitable analog for benchmarking reflection and propagation behavior on a public dataset.

\textbf{Symbol filtering for meaningful perturbation.} Not all RxTerms symbols are equally informative for benchmarking. Routes and dose forms (such as ``Mucosal'' or ``Injection 50 ml'') carry their meaning in the name itself, so an LLM-generated description adds little and a perturbation has nothing meaningful to corrupt. We therefore filter the pool before perturbation by scoring each symbol's name-informativeness on a 0 to 10 scale (rule-based for self-describing patterns, \texttt{gemini-2.5-flash} for the rest) and dropping symbols scoring 2 or higher. This eliminates 84\% of the pool, including every dose form and route. The 457 retained candidates are all ingredient names, opaque pharmaceutical terms such as ``metformin'' or ``tofersen'' whose meaning cannot be guessed from the name itself, more closely matching the industrial shorthands (e.g., \testdesc{GLS1}, \testdesc{OVRLY}) in our primary case study.

\textbf{Perturbation strategy.} On the filtered pool, we apply a \textit{deliberate perturbation} pipeline: \texttt{gpt-5-mini} produces the reference descriptions, and \texttt{gemini-2.5-flash} injects errors into them. Splitting these two roles between model families avoids the circularity of having one model both author and judge the ground truth. We synthesize 240 perturbed symbol descriptions covering three error categories (semantic confusion, granularity error, and surface edits), affecting 2,744 identifiers in total. For each scenario, we feed the perturbed-to-correct pair into RAID's reflective agent as a simulated expert edit and observe whether intent inference and propagation behave as expected.

\textbf{Metrics.} We report four metric groups. \textbf{Intent classification accuracy} measures the inference step. \textbf{Propagation precision}, \textbf{recall}, and \textbf{F1} are computed over the set of identifiers RAID proposes to update. For the revision step, \textbf{revision accuracy} is the fraction of identifier-level \texttt{apply}/\texttt{skip} decisions matching ground truth, and \textbf{cosine similarity} between generated and ground-truth descriptions captures semantic fidelity.

\textbf{Results.}
Table~\ref{tab:rxterms-results} summarizes the performance of identifier edit reflection across the three error types. Intent inference is essentially perfect: only 1 of the 240 edits is misclassified (a semantic confusion edit), and surface rephrases are correctly recognized as no-propagation cases 100\% of the time. Propagation retrieval achieves perfect recall (1.000) on every error type, with precision driven down by a set of substring-collision edits, which are analyzed below. 
Revision accuracy is high for semantic confusion (93.9\%) and granularity (87.9\%); the lower number for surface rephrase (49.7\%) is caused by a design choice, also discussed below. Cosine similarity remains around 89\% across types, indicating that the generated descriptions are semantically close to ground truth regardless of the apply/skip decision.

\begin{table}[h]
\caption{Symbol edit reflection, propagation, and revision results by error type (\%). Prop. = propagation; Rev. Acc. = revision accuracy; Cos. Sim. = cosine similarity.}
\label{tab:rxterms-results}
\resizebox{\columnwidth}{!}{%
\begin{tabular}{lcccccc}
\toprule
Error Type & Intent Acc. & Prop.\ P & Prop.\ R & Prop.\ F1 & Rev.\ Acc. & Cos.\ Sim. \\
\midrule
Sem.\ confusion  &  98.8 &  97.6 & 100.0 &  98.8 & 93.9 & 88.6 \\
Granularity     & 100.0 &  98.3 & 100.0 &  99.1 & 87.9 & 88.4 \\
Surface rephrase & 100.0 & 100.0 & 100.0 & 100.0 & 49.7 & 90.3 \\
\midrule
\textbf{Overall} & \textbf{99.6} & \textbf{98.6} & \textbf{100.0} & \textbf{99.3} & \textbf{76.3} & \textbf{89.1} \\
\bottomrule
\end{tabular}}
\end{table}

\textbf{Error analysis.} Of the 160 symbol edits that should propagate,\footnote{Of the 240 edits, the 160 semantic-confusion and granularity edits require propagation; the 80 surface edits do not.} one semantic-confusion edit was misclassified as surface during intent inference and therefore skipped, leaving 159 symbol edits that actually executed retrieval. These 159 edits cover 2,744 identifiers in total. Among the 159 edits, 153 produced zero false positives at the identifier level. The remaining six edits collectively produced 139 retrieval false positives, all caused by \textit{lexical substring confusion}: the edited symbol name appears as a substring inside pharmacologically distinct identifiers (e.g., a tretinoin edit retrieves isotretinoin identifiers; a rutin edit retrieves \textit{ibrutinib}, \textit{zanubrutinib}, and \textit{acalabrutinib} identifiers). This suggests that false-positive risk on this benchmark can be mitigated by improved retrieval methods. These retrieval errors then cascade into 115 false revisions, where the LLM regenerates already-correct descriptions for the falsely retrieved identifiers. Only one of the six FP-affected edits (ibrutinib) was caught downstream by the revision step's \texttt{apply}/\texttt{skip} filter, indicating that the revision stage offers limited but real defense in depth. Finally, the comparatively low surface-rephrase revision accuracy (49.7\%) is caused by a design choice: surface edits do not require propagation by default, but our formative study found that some experts still wanted the option to propagate them, so we did not specifically design instructions to suppress propagation in this case. The 90.3\% cosine similarity for surface rephrase confirms that the agent's outputs remain semantically faithful when propagation is applied.

\subsection{User Study}

We conducted a qualitative user study to assess the prototype's \textbf{technical feasibility} with proprietary data and \textbf{usability}, as well as to understand the user perceptions in real-world workflows.
Four SMEs (\textit{E1} to \textit{E4}), each with more than three years of semiconductor test analysis experience, performed curation tasks on a proprietary knowledge base, followed by a usability survey and semi-structured interview.
Each session began with a 10-minute walkthrough of the reflection-propagation algorithm. The SMEs then executed at least two corrections (e.g., redefining a test) and managed the resulting propagation, thinking aloud as they verbalized their reasoning steps. Afterward, they completed a short survey covering two UMUX-Lite questions and a question about intent-inference quality, followed by a semi-structured interview about their perceptions of the system.

The prototype received a mean UMUX-Lite score of \textbf{3.75} and a score of \textbf{4} (out of 5) for intent-inference quality, and SMEs reported that the agent's reflection mostly captured their intent, indicating the technical feasibility of the framework. Beyond these aggregate measures, several qualitative themes emerged.
 
\textbf{Emerged propagation strategies.} Experts were inspired by the cases they encountered in the study and proposed new propagation patterns. First, \textbf{chain-of-reflection}: when the agent infers that a symbol's definition needs updating, experts sometimes expected a second round of reasoning over symbol \emph{variants} (e.g., updating \texttt{GLS1} should prompt the agent to consider \texttt{GLS2} and \texttt{GLS3}). 
Second, \textbf{propagation along semantic rather than syntactic relations}: existing KB connections are predominantly syntactic, but experts pointed to entities connected through derivation, paired use, or shared physical context. Although this strategy is not supported by the current system because such connections are not recorded in the knowledge base, the suggestion provided insights for future work and for potential generalization of the propagation method.
 
\textbf{Implicit feature requests.} Three needs were articulated through user reactions rather than direct requests. (1) When SMEs disagreed with a proposed propagation, they asked whether they could correct it: the current interface enables them to adjust the search scope at \textit{Step 2 Search card} but not directly edit individual revisions. (2) Before clicking apply, SMEs expressed concern about how far the change would reach and which entries would be touched, but the interface only shows the number of affected identifiers in a list, which does not vividly demonstrate the propagation impact. (3) Even when proposals were correct, reviewing them one by one was time-consuming. 
Together these point to design implications we discuss in the next section.
 
\section{Discussion and Future Work}
\label{sec:discussion}

Our framework adapts the ``reflect-then-update'' pattern studied in agent memory research, where agents self-edit an external storage to accumulate experience over time~\cite{Packer2023-MemGPT, Park2023-Generative, Zhao2024-ExpeL, Hu2026-Memory-Age}. 
RAID differs in two ways: the storage being updated is a shared organizational KB rather than the agent's own memory, so the agent is not the principal but a human expert is; and the update path is mediated, with stepwise checkpoints between intent inference, retrieval, and revision.
 
\textbf{From identifier dictionaries to document knowledge bases.} Our case study presents one data structure of organization KB: each entry is a short description; and entries are linked through a shared symbol vocabulary, the symbol-to-identifier map defines which entries are candidates for propagation. 
Another common organizational data format is a corpus of long, hierarchical documents chunked and indexed for Retrieval-Augmented Generation (RAG)~\cite{Gao2023-RAG}. For chunks with primarily flat structure, a correction has no obvious place to land. Recent work already produces the structure that the reflect-and-propagate method needs: GraphRAG extracts entities and relationships into a typed knowledge graph~\cite{Edge2024-GraphRAG}; KG\textsuperscript{2}RAG links chunks through fact-level relationships drawn from an external KG~\cite{Zhu2025-KG-Guided-RAG}; hierarchical methods such as KohakuRAG and SF-RAG preserve the native document tree~\cite{Yeh2026-KohakuRAG, Yu2026-SF-RAG}. Our ``reflect-and-propagate'' method can utilize these processing techniques and retrieve relevant document pieces based on both structural and semantic connections.
 
\textbf{Editable propagation and agent self-improvement.} When a user rejects or rewrites a proposed revision, that signal carries information about what the agent's intent inference or retrieval missed, which fits into self-evolving agent frameworks~\cite{Gao2025-Self-Evolving}. Adapting them to RAID creates opportunities to study whether and how heuristics accumulated from one expert's session transfer to another's.
 
\textbf{Visualization as a tool for propagation scope.} A complementary direction is visual analytics representations of the KB and of the affected neighborhood, so that experts can judge propagation breadth before costly steps execute. The need grows when propagation moves from a few hundred linked identifiers to the larger, less regular neighborhoods of a chunked document corpus.

\section{Conclusion}

We presented RAID, a reflective agent system that scales a single expert edit into systematic updates across a compositional knowledge base.
By inferring the semantic intent behind an edit and propagating corrections through a chain-of-tools plan, RAID seeks to amplify limited expert feedback for maintaining large-scale identifier dictionaries.
Our evaluation provides initial evidence for the technical feasibility of reflective edit propagation and suggests the system's potential to help scale specialized expertise across industrial knowledge bases.

\bibliographystyle{ACM-Reference-Format}
\bibliography{bib/paperpile}

\clearpage
\newpage

\appendix
\section{System Design Details}

\subsection{Semiconductor Test Dictionary}
 
Our case study uses a semiconductor test dictionary, where each test is encoded as a compositional identifier assembled from reusable engineering shorthands. 
For example, \testdesc{RT7-LMP\_\_VOP-GLS1-FP12\_\_ALIG\_OVRLY} decomposes into symbol groups corresponding to a process module, a measurement parameter, and an operational condition. 
A single chip undergoes thousands of such tests during manufacturing, and the meaning of each shorthand typically lives in fragmented internal documentation or in the heads of the engineers who designed the naming conventions, limiting the generalizability of automated decoders, including those based on LLMs.
The initial KB descriptions used in our study are synthesized by an LLM agent that decomposes each identifier, infers symbol-level definitions, and assembles them into a full-text description. This baseline is functional but contains technical inaccuracies that motivate the expert-in-the-loop correction workflow described in the main text.\footnote{Considering corporate confidentiality while maintaining technical depth, we use synthesized data throughout this paper that mimics the structure of real-world semiconductor test sequences.}

\subsection{Formative Study}
\label{appendix:formative-study}

To inform the system design, we conducted a formative study with six semiconductor subject matter experts (SMEs): two through individual interviews and four through a focus-group workshop. Participants performed audits on LLM-generated entries via a dashboard prototype, after which we presented the reflective agent concept and elicited expectations about what aspects of an edit the agent should reason over.
 
The study revealed three findings that shaped the design: 
(1)~experts' corrections often target a specific symbol's underlying definition rather than only the edited identifier, implying that a single edit should propagate to structurally related entries; 
(2)~experts prefer to make corrections on the assembled identifier description rather than on individual symbol metadata, as the assembled form aligns with their diagnostic workflow; 
and (3)~experts welcomed the propagation concept but were cautious about over-propagation across large datasets based on a single inferred intent. 

We summarized three design requirements.

(1)~\textbf{Expert-in-the-loop validation}: experts must remain the final authority. No changes are committed to the KB without explicit human approval.
(2)~\textbf{Explainability}: the system must surface its reasoning, including how and why a specific edit was linked to other records, so that experts can detect potential misinterpretations.
(3)~\textbf{Flexible intervention}: experts should be able to review and modify the proposed propagation logic before any change is applied.

\subsection{Reflection Agent Prompt}
We present the identifier edit reflection prompt in Figures~\ref{fig:reflection-prompt-1} and~\ref{fig:reflection-prompt-2}.

\begin{figure*}[htp]
\par\noindent
\fcolorbox{black!50}{gray!15}{%
  \makebox[\dimexpr\linewidth-2\fboxsep-2\fboxrule\relax][l]{%
    \strut\hspace{2mm}\textbf{Reflection agent prompt}%
  }%
}%
\par\nointerlineskip\vspace{-\fboxrule}%
\begin{lstlisting}[
    basicstyle=\ttfamily\small,
    breaklines=true,
    basewidth=0.5em,
    columns=fullflexible,
    lineskip=-1pt,
    frame=single,
    framerule=0.4pt,
    rulecolor=\color{black!50},
    backgroundcolor=\color{white},
    framesep=2mm,
    xleftmargin=2mm,
    xrightmargin=2mm,
    framexleftmargin=0mm,
    framexrightmargin=0mm,
    aboveskip=0pt,
    belowskip=0pt,
]
Your job is to infer expert's reasoning behind the correction and reflect on whether the change suggests an update to the meaning of any symbol connected to the identifier.
Return with steps "update_description", "filter", and "batch_llm_revise_description" (must follow this order).

## ACTION PARAMETERS
In the response json "steps" field, you need to propose a list of actions in the order of "update_description", "filter", "batch_llm_revise_description" to demonstrate your reflection on user's edits.
Besides the required fields "action", "target", and "title", each action also contains a "params" field with additional parameters.
Below are the details of each type of actions's parameters.
ACTION 1: update_description
Description: Update the description of a symbol or test. Set "target" as "identifier" if you want to update test (identifier) description, otherwise set as "symbol".

**Parameters**:
- name: test name or symbol name
- old_description: the old description before the change
- new_description: the new description to set
- reasoning: reasoning for the change
ACTION 2: filter
Description: This action is to filter identifiers or symbols. The parameters are used to filter the items.

**Parameters**:
- nl_query: natural language query to filter items. The system will convert the nl_query into proper filtering conditions. Note that the searched items should be in the same database as the source item, if `database` is available in the item's `metadata`.
ACTION 3: batch_llm_revise_description
Description: This action will trigger the system to revise existing description for a batch of identifiers or symbols based on an instruction.

**Parameters**:
- instruction: The instruction is sent to LLM to guide the description proposal.
- relevant_symbols (Optional): A list of symbol names that are relevant to this action.
- relevant_identifiers (Optional): A list of identifier names that are relevant to this action.

## EXAMPLES
{examples}

\end{lstlisting}
\caption{Reflection agent prompt template. The prompt instructs the LLM to infer expert intent from an edit and plan propagation via a sequence of tool calls.}
\label{fig:reflection-prompt-1}
\end{figure*}

\begin{figure*}[htp]
\par\noindent
\fcolorbox{black!50}{gray!15}{%
  \makebox[\dimexpr\linewidth-2\fboxsep-2\fboxrule\relax][l]{%
    \strut\hspace{2mm}\textbf{Reflection agent prompt (continued)}%
  }%
}%
\par\nointerlineskip\vspace{-\fboxrule}%
\begin{lstlisting}[
    basicstyle=\ttfamily\small,
    breaklines=true,
    basewidth=0.5em,
    columns=fullflexible,
    lineskip=-1pt,
    frame=single,
    framerule=0.4pt,
    rulecolor=\color{black!50},
    backgroundcolor=\color{white},
    framesep=2mm,
    xleftmargin=2mm,
    xrightmargin=2mm,
    framexleftmargin=0mm,
    framexrightmargin=0mm,
    aboveskip=0pt,
    belowskip=0pt,
]
## TARGET IDENTIFIER
- Name: {target_identifier__name}
- Description: {target_identifier__description}
- Description Source: {target_identifier__description_source}
- Metadata: {target_identifier__metadata}
- Status: {target_identifier__status}
- Human Validated: {target_identifier__human_validated}
- Last Updated At: {target_identifier__updated_at}

## CONNECTED SYMBOLS
<symbol_1_name>
- Symbol name: {name}
- Description: {description}
- Description Source: {description_source}
- Status: {status}
- Relationship to Identifier: {relationship}
</symbol_1_name>
<symbol_2_name>
...
</symbol_2_name>
(one block per connected symbol, rendered from symbol_entity_template)

## EXPERT'S CORRECTION
- Original description: {old_description}
- Corrected description: {new_description}

## INSTRUCTION
Analyze the change from the original description to the expert's correction.
1. **Identify the core difference**: What specific pharmacological fact was changed? Was it the mechanism of action, therapeutic class, clinical distinction, dose form characterization, or route description?
2. **Attribute the change to a symbol**: Which component symbol (ingredient, dose form, or route) in the connected symbols most likely corresponds to the changed part of the description? Use the CONNECTED SYMBOLS section to identify the source.
3. **Assess semantic impact**: Use the criteria below to classify the change.
4. **Formulate a propagation plan**: If the correction reveals that a symbol's description needs update, propose in sequence:
- An update_description action for the affected symbol with the corrected meaning;
- A filter action that search in the database for the affected identifiers;
- A batch_llm_revise_description action for sibling identifiers that share the same symbol

If the correction is only a surface rephrase, set suggest_propagation to false but still propose a plan with full steps.

{format_instructions}

\end{lstlisting}
\caption{Reflection agent prompt template (continued). The prompt provides the target identifier context, connected symbols, the expert's correction, and instructions for classifying the edit as semantic or surface.}
\label{fig:reflection-prompt-2}
\end{figure*}



\section{Evaluation Details}
\label{appendix:evaluation}

\subsection{RxTerms Benchmark Construction}
\label{appendix:benchmark}

Section~\ref{sec:quant-eval} describes the symbol filtering and perturbation strategy. Here we detail the three error types and the sampling procedure. 

We define three error types: 

(1)~\textbf{Semantic confusion}: the symbol description is replaced with that of a related but distinct compound (e.g., cholecalciferol described as ergocalciferol). Confusion pairs are drawn from LLM-generated \textit{variation groups} (families of compounds related by salt form, ester, or narrow pharmacological subclass). 
(2)~\textbf{Granularity error}: the symbol description is replaced with a vague, class-level description that omits distinguishing clinical details (e.g., \textit{``A dihydropyridine calcium channel blocker that relaxes vascular smooth muscle''} replaced with \textit{``A blood pressure medication''}). 
(3)~\textbf{Surface edits}: cosmetic changes (case normalization, synonym substitution) that alter wording without changing meaning. These do \textit{not} need propagation. 

    
    

\textbf{Sampling.} The 457 retained ingredient symbols are deterministically split into three groups (one group per error type) using a fixed random seed. From each group, we draw a stratified sample of 80 symbols, stratified by \textit{propagation count} (the number of identifiers linked to each symbol) into low (1 to 3), medium (4 to 10), and high (11+) bins with proportional allocation. The resulting 240 sampled symbols yield 160 propagating edits (80 semantic confusion + 80 granularity) and 80 non-propagating surface edits, affecting 2,744 identifiers in total. Each propagating edit has a known ground-truth set of affected identifiers derived from the compositional mapping between symbols and identifiers.




\subsection{Error Analysis Details}
\label{appendix:error-analysis}

\paragraph{Lexical substring confusion (retrieval false positives).}
All 139 retrieval false positives originate from six edits whose edited symbol name appears as a substring inside pharmacologically distinct identifiers. Table~\ref{tab:fp-breakdown} lists the affected pairs.

\begin{table}[h]
\caption{Retrieval false positives caused by lexical substring confusion. The edited symbol name appears as a substring of unrelated identifiers, which the NL search step over-retrieves.}
\label{tab:fp-breakdown}
\resizebox{\columnwidth}{!}{%
\begin{tabular}{llcl}
\toprule
Edited symbol & FP identifiers & FPs & Error type \\
\midrule
tretinoin        & iso\textbf{tretinoin}                                      & 43 & granularity \\
tetrabenazine    & deu\textbf{tetrabenazine}, val\textbf{benazine}            & 30 & granularity \\
dextroamphetamine & lis\textbf{dexamfetamine}                                 & 28 & sem.\ confusion \\
rutin            & ib\textbf{rutin}ib, zanub\textbf{rutin}ib, acalab\textbf{rutin}ib & 28 & sem.\ confusion \\
citalopram       & es\textbf{citalopram}                                      &  8 & sem.\ confusion \\
ibrutinib        & remi\textbf{brutinib}                                      &  2 & sem.\ confusion \\
\bottomrule
\end{tabular}}
\end{table}

\paragraph{Cascading false revisions.}
Five of the six retrieval-FP edits propagate downstream into 115 false revisions, where the agent regenerates already-correct descriptions on the falsely retrieved identifiers. 
Table~\ref{tab:false-applies} summarizes the cascade.

\begin{table}[h]
\caption{
False revisions cascading from the retrieval FP. 
}
\label{tab:false-applies}
\resizebox{\columnwidth}{!}{%
\begin{tabular}{lc}
\toprule
Edit & False applies \\
\midrule
tretinoin $\rightarrow$ isotretinoin identifiers              & 42 \\
tetrabenazine $\rightarrow$ deutetrabenazine identifiers      & 30 \\
dextroamphetamine $\rightarrow$ lisdexamfetamine identifiers  & 29 \\
citalopram $\rightarrow$ escitalopram identifiers             &  8 \\
rutin $\rightarrow$ *brutinib identifiers                     &  6 \\
ibrutinib $\rightarrow$ remibrutinib identifiers (intercepted) &  0 \\
\bottomrule
\end{tabular}}
\end{table}

\end{document}